Analysis of cellular responses of macrophages to zinc ions and zinc oxide nanoparticles: a combined targeted and proteomic approach

Sarah Triboulet [1, 2, 3], Catherine Aude-Garcia [1, 2, 3], Lucie Armand [4], Adèle Gerdil [5], Hélène Diemer [6], Fabienne Proamer [7], Véronique Collin-Faure [1, 2, 3], Aurélie Habert [5], Jean-Marc Strub [6], Daniel Hanau [7], Nathalie Herlin [5], Marie Carrière [4], Alain Van Dorsselaer [5], Thierry Rabilloud [1, 2, 3]*

1: Univ. Grenoble Alpes, Laboratory of Chemistry and Biology of Metals, Grenoble, France

2: CEA Grenoble, iRTSV/CBM, Laboratory of Chemistry and Biology of Metals, Grenoble, France

3: CNRS UMR 5249, Laboratory of Chemistry and Biology of Metals, Grenoble, France

4: UMR E3 CEA-Université Joseph Fourier, Service de Chimie Inorganique et Biologique, Laboratoire Lésions des Acides Nucléiques (LAN), Grenoble

5: CEA-DSM, IRAMIS, SPAM-LFP (URA CEA CNRS 2453), Saclay

6: Laboratoire de Spectrométrie de Masse BioOrganique (LSMBO), Université de Strasbourg, IPHC, 25 rue Becquerel 67087 Strasbourg, France. CNRS, UMR7178, 67037 Strasbourg, France.

7: UMR_S949, INSERM-UdS, EFS-Alsace, 10, rue Spielmann, 67065 Strasbourg

*: to whom correspondence should be addressed:
Laboratoire de Chimie et Biologie des Métaux, UMR CNRS-CEA-UJF 5249, iRTSV/LCBM, CEA Grenoble, 17 rue des martyrs, F-38054 Grenoble Cedex 9, France

thierry.rabilloud@cea.fr

ABSTRACT

Two different zinc oxide nanoparticles, as well as zinc ion, are used to study the cellular responses of the RAW264 macrophage cell line. A proteomic screen is used to provide a wide view of the molecular effects of zinc, and the most prominent results are cross-validated by targeted studies. Furthermore, the alteration of important macrophage functions (e.g. phagocytosis) by zinc is also investigated. The intracellular dissolution/uptake of zinc is also studied to further characterize zinc toxicity. Zinc oxide nanoparticles dissolve readily in the cells, leading to high intracellular zinc concentrations, mostly as protein-bound zinc. The proteomic screen reveals a rather weak response in the oxidative stress response pathway, but a strong response both in the central metabolism and in the proteasomal protein degradation pathway. Targeted experiments confirm that carbohydrate catabolism and proteasome are critical determinants of sensitivity to zinc, which also induces DNA damage. Conversely, glutathione levels and phagocytosis appear unaffected at moderately toxic zinc concentrations.

INTRODUCTION

Zinc oxide nanoparticles are among the most widely used nanoparticles. Apart from their possible role in nanomedicine (e.g. in [1]), they are also widely used in sunscreens [2] or as antibacterials [3]. However, zinc oxide nanoparticles are known to be toxic in man, as they are for example one of the causal agents of the metal fume fever [4]. There are also several reports on zinc oxide toxicity on different cellular models [5-11]. Although some reports show a clear role for the nanoparticles themselves [7,12], several reports demonstrate a clear role of dissolved zinc ion in the toxicity of zinc oxide nanoparticles [13-16]. As to the role of dissolved zinc ion, some reports suggest a major role for extracellular dissolution [11,15,17], while other reports suggest that intracellular dissolution in lysosomes after uptake of nanoparticles is predominant [7, 12, 13, 18].

As to the toxicity mechanisms of zinc oxide nanoparticles, most of the studies imply the oxidative stress pathway[5, 9, 19], leading to DNA damage [5, 8], although interference with ion homeostasis has also been implicated [9,20]. However, these are mostly targeted studies investigating classical mechanisms in nanoparticles toxicity, and studies using wider and less oriented approaches are still rare [21].

Due to their known implication in diseases related to particulate matter such as asbestosis [22, 23], and the well-known inflammatory dimension of the zinc oxide-induced metal fume fever [24,25], myeloid cells and especially macrophages are a typical target cell type when investigating the toxicity of zinc oxide nanoparticles. Thus, myeloid cells have been used in several studies dealing with zinc oxide toxicity [6, 11, 21, 26, 27]. Most of these studies have concentrated on the classical chronic inflammation scheme, and investigated the production of pro-inflammatory cytokines upon treatment with zinc oxide. The results are however conflicting, some studies describing the production of some cytokines [26,27] while others do not describe any cytokine production [6].

However, a wide scope approach aiming at finding the key responses of macrophages to zinc oxide nanoparticles is still missing, despite the recent transcriptomic work recently described [21]. This is why we undertook a combined targeted and proteomic study of the responses of the mouse macrophage cell line RAW264 to zinc ion and to two different types of zinc oxide nanoparticles (coated and uncoated). RAW264 cells are one of the classical macrophage models and have been often used for studying the effects of nanoparticles in macrophages (e.g. in [26, 28-33]. They are also one of the few cell lines able to produce NO upon LPS stimulation [34], a feature absent from the other classical human macrophage model THP-1. The rationale of this comparison was to check if the coating can modify not only the toxicity of the nanoparticle [19], but can also alter the cellular responses to the nanoparticles.

RESULTS

Nanoparticles characterization

The two commercial nanoparticles were characterized by several methods, and the results are summarized in Table 1 and in supplementary figures 1 to 3. The zinc oxide nanoparticles were rod shaped, while the cationic coated zinc oxide was more spherical. The XRD diagrams conformed to those of nominal zinc oxide, showing no crystalline impurities. The zeta potential was positive for both nanoparticles, with the cationic coating accounting for an increase in the positive surface charge. Finally, the nanoparticles could be dispersed in complete culture medium as aggregates of ca. 200 nM in diameter.

In order to verify that the nanoparticles were not degraded prior to their contact with the cells, we checked their dissolution in the 50mg/ml intermediate nanoparticules dispersion (corresponding to 625mM zinc). Only 1mM dissolved zinc was measured after 7 days.

Determination of the effective doses

For carrying out a proteomic study, we had to determine the dose for which we would obtain the best compromise between viability and biological effect. We therefore decided to use a LD20, i.e. a dose leading to a cell mortality of 20%, knowing that the mortality of a control culture is around 5%.

The LD20 was obtained at 8 µg/ml for the ZnO nanoparticles, 7µg/ml for the coated cationic ZnO nanoparticles, and at 120µM for zinc ion (supplementary figure 4). These concentrations were used for the subsequent studies.

Determination of nanoparticles fate in cells

Transmission electron microscopy was performed to determine the fate of the nanoparticles when in presence of macrophages. These experiments (Figure 1) showed that nanoparticles were present in vacuolar structures identified as multivesicular bodies and as lysosomes. These results were in line with what can be expected from professional phagocytes. Moreover, these results demonstrate that macrophages seem to ingest both nominal nanoparticles and nanoparticles aggregates present in the culture medium and to break them down in the endosomes-lysosomes. Indeed, eroded nanoparticles can easily be seen (Figure 1B and 1C), demonstrating this intracellular erosion phenomenon.

In order to gain further insights into the intracellular zinc concentration, we measured the zinc concentration that could be released from a post nuclear supernatant prepared from cells exposed to zinc ion or ZnO nanoparticles. The results, shown in Table 2, demonstrate that zinc is concentrated in cells compared to the original input. For example, the values obtained with zinc ion convert to an intracellular concentration of 187± 12.5 µM, while the input is only 100µM zinc in the culture medium. These values are orders of magnitude higher than those recently published for the concentrations of free zinc ion after exposure to zinc salts and ZnO nanoparticles [20]. It should be emphasized, however, that the two studies do not measure the same parameter. The study using zinc-sensitive fluorescence probes [20] measures only free zinc, and thus does not measure zinc complexed to macromolecules (e.g. proteins) whereas our study implements a strong protein denaturation step to release protein-bound zinc and to measure all intracellular zinc ion. This demonstrates in turn that most intracellular zinc is not free but protein-bound, which prompted us to analyze zinc-binding proteins.

Proteomic analyses

Two sets of proteomic analyses were performed. The first one dealt with the quantitative analysis of the proteome on whole cell extracts prepared from control cells vs. cells exposed to zinc ion or to both types of ZnO nanoparticles. This analysis pointed to the cellular responses to the presence of excess zinc in terms of protein quantitative changes. The second type dealt with zinc-binding proteins, providing a list of putative targets in case of a zinc overload. This list does not necessarily superimpose to the list of zinc-containing proteins under normal zinc load. As the extract used in the affinity enrichment is prepared under normal growth conditions, it is highly likely that the bona fide zinc binding proteins will be already loaded with zinc and will not bind to the extra zinc ion presented via the affinity column. Conversely, we expect to retain on the column the proteins that do not necessarily bind zinc under normal conditions, but that can bind it when present in excess, i.e. in the conditions prevailing during zinc overload.

The proteomic analysis of whole cell extracts is detailed in Figure 2 and Table 3 . With 2780 reproducibly detected spots, and using an average of 3 spots per protein [35] this proteomic screen probed the proteome to a depth of ca 900 gene products, i.e. ca.15% of the total proteome. Despite this rather limited depth, we could detect reproducible modulations forproteins belonging to several functional classes. The median coefficient of variations of spots was 28%, i.e. marginally higher than in typical 2D DIGE experiments, where coefficient of variations range from 18 to 28%, depending on the sample [36-39]. The protein changes were detected through the use of a variance-based screen [40]. Compared to a fold-change screen, this process compensates automatically for the variance of each spot. This excludes automatically spots with a high coefficient of variation, but enables to take into account small but reproducible changes when the coefficient of variation is low, thus avoiding the arbitrary exclusion of changes that can be biologically meaningful [40]. As an example, the endoplamic reticulum protein ERP29 is selected as significantly modulated, although the fold change is only 0.8 fold. This is due to the fact that the coefficient of variation of this spot is only 8% in the control group and 4% in the zinc ion-treated group.

Metabolic enzymes, cytoskeletal proteins and proteins implicated in the proteasomal degradation are strongly represented in the list of modulated proteins. Moreover, more than a third of these modulated proteins (17/45) are significantly changed only by the zinc ion, but not by the nanoparticles. Very few proteins (4) are significantly changed by the three treatments, and very few are specific of one type of nanoparticles (4 for zinc oxide and 3 for cationic zinc oxide). It is also interesting to compare our proteomic results to the transcriptomic results recently published on immune cells and zinc oxide [21]. Only two proteins (ERP29, spot F1, and major vault protein, spot C8) are found in common between our list of modulated proteins and the list of modulated mRNAs in macrophages, an additional four proteins being found in common with modulated RNAs in Jurkat cells (ADSS spot N2 , DLAT spot M3, GCLM spot O2,  and calcyclin-binding protein spot D2).

The proteomic analysis of zinc binding proteins is detailed in Figure 3 and Table 4. First, it should be noted that the background given by an "empty" resin (i.e. without a bound metal) was extremely low, as a basically empty 2D gel was obtained (data not shown). However, to minimize any possible bias, the experiments were carried out on three different biological replicates and only the proteins significantly enriched in the zinc column compared to the starting extract ($p \leq 0.05$) were further analyzed and listed in Table 2. Here again, metabolic enzymes and cytoskeletal proteins were strongly represented. The two tables show a low overlap, however, pointing out the complementarity of the two analyses.

Validation studies

The inclusion of small but reproducible protein changes means in turn that these changes cannot be validated easily  by classical biochemical techniques at the protein expression level. For example, protein blotting often shows a technical variability well above 20%, and  a response curve often lower than the one of 2D electrophoresis (e.g. in[38]), making this technique unsuitable for the validation of small fold changes. This renders functional validation even more necessary [41], to confirm the biological relevance of the proteomically detected protein modulations.

Characterization of the oxidative stress response

Oxidative stress is now used as a toxicological marker for almost all toxicants, and zinc ion and zinc oxide are no exception to this rule [5, 14]. However, we did not detect any change in the classical oxidative stress response proteins, as opposed to what was observed with copper oxide nanoparticles [42]. Nonetheless, we detected an increase in two non classical proteins that can be involved in the antioxidant response, namely flavin reductase (spot O3) and glutamate cysteine ligase (GCLM, spot O2). The latter protein is of interest, as it controls the limiting step in glutathione biosynthesis. Furthermore, glutathione depletion has been suggested as an important mechanism in zinc toxicity [43]. We therefore measured the intracellular levels of reduced glutathione by a monochlorobimane conjugation approach [44]. The results, shown on Figure 4A, demonstrate a moderate but reproducible increase in GSH levels, validating the increase of GCLM detected via proteomics (fold change 1.6 to 1.8 in Table 3). This suggests that macrophages are able to counteract the GSH chelation due to zinc ion by de novo synthesis of glutathione, a situation quite different from the one observed in neurons [43].

Another interesting protein detected via our proteomic screen and involved in the oxidative stress response is the flavin/biliverdin reductase (spot O3), which reduces biliverdin into bilirubin. This bilirubin/biliverdin cycle has been implied in hydrogen peroxide destruction [45]. As this protein has a detectable enzymatic activity, we decided to confirm the proteomic result by a direct assay of the biliverdin reductase activity (Table 5). A significant induction was observed, as detected by the proteomic screen.

Characterization of the phagocytic response

As we observed several changes in proteins involved in the actin (e.g. ARP proteins spots C5 and C9, lsp1 protein spot C3, glia maturation factor spot C1, twinfilin spot C4 or gelsolin spots C7) and tubulin cytoskeleton (MAPRE1 spot C2 and TCP1 spot C6), we could anticipate perturbations in the phagocytic capacity of zinc-treated macrophages. We thus assessed this important function of macrophages upon treatment with zinc ion or ZnO nanoparticles. The results, presented on Figure 4B, show that macrophages are still highly phagocytic even when treated with zinc, as long as the concentrations remain non toxic.

DNA damage

As several nanoparticles have been shown to induce DNA damage [8, 46-48], we decided to test whether zinc-based nanoparticles can induce DNA damage, using an alkaline comet assay. The results, shown on Figure 5, demonstrate that zinc ion, zinc oxide but not cationized zinc oxide, are able to induce DNA damage even at the moderately toxic LD20 dose. This also confirms previous results , e.g. on normal nasal mucosa cells [8].

Metabolic perturbations

Our proteomic screens highlighted several proteins implicated in energy metabolism, and more specifically in the glycolysis and pentose phosphate pathways (spots E1 to E5). These proteins were found in both screens, i.e. proteins responding to the zinc stress and zinc-binding proteins. In fact, several important energy metabolism enzymes have been shown to bind zinc and be inhibited by it [49-51].

We proceeded to a two step validation. First, the enolase and 6-phosphogluconate dehydrogenase activities were measured (Table 5). Here again, the enzymatic activities confirmed the proteomic results, showing an increase upon treatment of the cells with zinc. As the 6-phosphogluconate dehydrogenase was also found in the zinc binding proteins, we hypothesized that it could be inhibited by zinc in the context of a complete cell extract, and not only as a purified protein [52]. We therefore assayed the 6-phosphogluconate dehydrogenase activity in the presence of added zinc acetate, and found an IC50 close to 12µM. As this concentration is much lower than the one detected in our cell extracts upon zinc treatment, it can be reasonably hypothesized that the pentose phosphate pathway is impaired during zinc overload. This could contribute to the observed oxidative stress, as the pentose phosphate pathway is a major source of NADPH, required by many reductases implied in the defense against oxidative stress (e.g. glutathione reductase, biliverdin reductase and thioredoxin reductase). In contrast, GAPDH was marginally inhibited by zinc, with only 30% inhibition at 100µM zinc (data not shown).

Finally, we tested the final metabolite of the pentose phosphate and glycolysis pathways, i.e. pyruvate. First, zinc stress induced a higher level of pyruvate in cells, as shown in Figure 6A. This is in line with the activation of the glycolytic metabolism observed in occupational medicine [53], and suggested that pyruvate is a critical metabolite for survival in presence of a zinc stress, as in neuronal cells [54, 55]. Our results demonstrate that pyruvate acts as a survival factor upon zinc stress, both for zinc ion (Figure 6B), zinc oxide (Figure 6C) and cationized zinc oxide (Figure 6D). Thus, energy metabolism is one of the crucial determinants in resistance to zinc.

Our proteomic screens also highlighted several proteins of the proteasomal degradation pathway such as proteasome subunits (spots D4 and D6) and proteins implicated in the ubiquitination step (spots D5 and D1 to D3). In order to investigate the putative implication of this pathway in the response to zinc, we assessed the survival to zinc treatment after proteasome inhibition with the inhibitor MG132. The results, shown on Figure 7, demonstrate that proteasome inhibition increases zinc toxicity, which can be attributed to a synergistic effect between the inhibition of the proteasome by MG132 and the known inhibition of the proteasome by zinc [56].

DISCUSSION

When taken together, the results of both the proteomic and targeted experiments produce a landscape of the cellular response to zinc. Opposite to what has been observed with copper [42], zinc ion and zinc oxide do not induce by themselves a strong oxidative stress. This is consistent with the fact that zinc, unlike copper, iron or manganese, cannot undergo redox cycles between ions of different valency. Thus, the almost generic oxidative stress that has been observed with zinc oxide [5,14], can be attributed to a more general perturbation leading to an imbalance in cellular redox cycles and thus to ROS production.

From our results, a key determinant in zinc toxicity appears to be a general metabolic perturbation. Many carbohydrate catabolic enzymes are either induced upon zinc treatment (Table 1) or bind to zinc (Table 2), and the overall result is both an increase in pyruvate production and the evidence that pyruvate is a critical metabolite for cell survival during zinc treatment. This is in line with previous studies on neuronal cells [51], and suggests that an important determinant in zinc toxicity is a metabolic impairment leading to an "energy shortage" in zinc-treated cells, leading eventually to cell death.

Along a closely related line, the number of proteins able to bind zinc ions and the importance of a correct proteasome function also suggests an unfolded protein response. Indeed, if taking an average Mw of 50,000 for proteins, a 5mg/ml solution of proteins represents a 100µM protein concentration, to be compared with the 30µM zinc detected in these extracts upon zinc overload. Consequently, many proteins binding zinc when present in excess may become ill-folded due to zinc binding, and therefore degraded through the proteasomal pathway. This would explain in turn why proteasome function is also an important parameter in zinc tolerance. This role of the proteasome may also explain why zinc, acting as a proteasome inhibitor [56] also synergizes with prooxidants [57], which also induce a strong unfolded protein response.

Furthermore, our results also shed light on the controversy between the roles of extracellular and intracellular dissolution of the zinc oxide nanoparticles to explain the toxic mechanisms observed. The responses to the zinc oxide nanoparticles are included in the responses to zinc ion, which induces further specific responses. This supports the hypothesis that the main determinant of zinc oxide toxicity is the zinc ion resulting from nanoparticles dissolution. However, the fact that zinc oxide nanoparticles are still visible in the cells after 24 hours of treatment suggests that uptake of nanoparticles and intracellular dissolution within the cells may be a key determinant of zinc toxicity, at least in macrophages. The situation may be indeed quite different in cells with a very low endocytic/phagocytic activity, such as T lymphocytes [17].

Last but not least, when comparing in detail the results obtained on copper oxide [42] and on zinc oxide in the RAW264 cell line, very few common features emerge, although both types of nanoparticles show close toxicity. This clearly demonstrates, if further needed, that each nanoparticle will induce different toxic mechanisms and different cell responses. In this context, zinc seems to act as a general toxicant, acting on central cellular pathways, so that cells are metabolically impaired but still functional as long as they are alive. Conversely, copper seems to act as a more subtle toxicant, leading to cells that are still alive but functionally impaired.

CONCLUSIONS

When trying to understand the toxicity of zinc nanoparticles in macrophages, the fact that nanoparticles are detected in the cells argues in favor of a model where the nanoparticles are first internalized, then intracellularly dissolved, releasing zinc ion that is the real toxicant. As cells seem to be relatively permeable to zinc, the molecular effects of the nanoparticles can also be reproduced with zinc ion, even when dealing with large scale screens such as proteomic analyses. Our analyses also show that zinc toxicity is produced by a strong uptake of zinc in a protein-bound form up to sub-micromolar concentrations, leading to impairment of central metabolism and thus to energy shortage in the cell, and also very likely to an unfolded protein response. As proteasome is also inhibited by zinc, the proteasome pathway becomes an important determinant of zinc resistance.

EXPERIMENTAL SECTION

Most of the experiments were performed essentially as previously described [42] and the detailed procedures can be found in the supplementary material and methods

Nanoparticles

Zinc oxide nanoparticles, coated or not, were purchased from Sigma-Aldrich (catalog numbers 721077 and 677450 respectively). The aminopropyl silane coated particles, representing a product used in solar screens, were used as purchased. The uncoated zinc oxide nanoparticles were dispersed in water as a 5% w/v suspension by sonication for 60 minutes in a cup-horn instrument (BioBlock Scientific, Strasbourg, France) , under a 5°C thermostated water circulation. The actual size of the particles was determined after dilution in water or in complete culture medium by dynamic light scattering, using a Wyatt Dynapro Nanostar instrument. A Malvern HS 3000 instrument was also used to determine the zeta potential. The crystalline phases were identified using a powder diffractometer from Bruker and the Match software (crystal impact). The morphology of samples was observed by SEM (Scanning Electron Microscopy) and TEM (Transmission Electron Microscopy) after dispersion of the nanoparticles in water using an ultrasonic bath and deposition of a droplet on 200 mesh carbon lacey grids. The weight loss (from about 10mg of sample) was evaluated from annealing under air using a Thermogravimetric Analysis device (Setaram). The temperature cycle was heating at a rate of 10°C/min up to 1000°C followed by a dwelling time of 30 minutes and natural cooling.

To measure nanoparticles dissolution in the 50 mg/ml intermediate dispersion used prior to addition in the cell cultures, the dispersion was centrifuged at 270,000g for 45 minutes to sediment the nanoparticles [58].

The dissolved zinc concentration in the supernatant was then determined, after the appropriate dilution in water, by a xylenol orange colorimetric assay [59], using 0.1M MES buffer pH 6 and 60µM xylenol orange as an indicator.

Nanoparticles dissolution in cells

For the studies on cell extracts, the cells were treated with zinc oxide nanoparticles for 24 hours prior to harvest. The cells were then recovered by scraping, collected by centrifugation, washed three times with PBS, and the volume of the cell pellet was determined. A cell lysis solution (Hepes 20 mM pH 7.5, MgCl2 2 mM, KCl 50 mM, tetradecyldimethylammonio propane sulfonate (SB 3-14) 0.15% w/v) was added at a ratio of 5 volumes of solution per volume of cell pellet, and the cells were left to lyse on ice for 20 minutes. The suspension was then centrifuged at 1000g for 5 minutes, the supernatant collected, and recentrifuged at 270,000g for 45 minutes to sediment the nanoparticles [58].

Zinc concentration was then determined using an adapted xylenol orange colorimetric assay [59]. Briefly, 1 ml of medium/cell extract was first cooled in an ice bath for 30 minutes. TCA was added to a 5% w/v final concentration, and the mixture was left for 1 hour on ice to denature the proteins and release complexed zinc ion. The mixture was centrifuged for 10 minutes at 15,000g, the supernatant recovered and its volume measured. A neutralizing solution (2 M MES-Na salt) was added (0.25 volume/volume of supernatant) was added, followed by 0.03 volume of a 2 mM xylenol orange solution in water. The color was left to develop for 5 minutes and the absorbance measured at 565 nm. Under these conditions, no interference was observed with magnesium, calcium or iron. As the extract is diluted 5fold compared to the initial cell volume, the concentration read in the assay is multiplied by 5 to obtain the intracellular zinc concentration.

Cell culture

The mouse macrophage cell line RAW 264.7 was obtained from the European Cell Culture Collection (Salisbury, UK). The cells were cultured in RPMI 1640 medium + 10% fetal bovine serum. Cells were seeded at 200,000 cells/ml and harvested at 1,000,000 cells per ml. For treatment with zinc ion or zinc oxide nanoparticles, cells were seeded at 500,000 cells/ml. They were treated with zinc on the following day and harvested after a further 24 hours in culture. Cell viability was measured by the neutral red uptake assay [60]. All experiments were carried out at least in triplicate on independent cultures.

Phagocytosis activity measurement

The phagocytic activity was measured after treatment with zinc using fluorescent latex beads (1μm diameter, green labelled, catalog number L4655 from Sigma) and flow cytometry, essentially as described in [61,42].

Transmission Electron Microscopy

Transmisson electron microscopy was carried out as previously described [42] on cells fixed with glutaraldehyde, included in Epon and post stained with lead citrate and uranyl acetate. Energy dispersive X-ray (EDX) analysis was performed on sample sections. Spectra were obtained using a large angle SDD-EDS attached to a Jeol 2010 TEM.

Intracellular glutathione measurements

Intracellular glutathione levels were analyzed by the monochlorobimane technique [44], with some modifications [62,42].

Pyruvate assay

The pyruvate content of cells was determined using the pyruvate assay kit from Sigma (ref # MAK071), used according to the manufacturer instructions.

Enzyme assays

The enzymes were assayed according to published procedures. Enolase was assayed by the direct conversion of 2-phosphoglycerate into phosphoenolpyruvate, detected at 240nm [63]. 6-phosphogluconate dehydrogenase was assayed by a coupled assay using Nitro blue tetrazolium as the final acceptor and phenazine methosulfate as a relay [64]. Biliverdin reductase was assayed directly for the NADPH-dependent conversion of biliverdin into bilirubin, followed at 450nm [45].

The cell extracts for enzyme assays were prepared by lysing the cells for 20 minutes at 0°C in Hepes 20 mM pH 7.5, $MgCl_2$ 2 mM, KCl 50 mM, EGTA 1 mM, SB 3-14 0.15% w/v, followed by centrifugation at 15,000g for 15 minutes to clear the extract. The protein concentration was determined by a dye-binding assay [65].

For the zinc inhibition studies, the cell extracts were first diluted in the assay buffer (final protein concentration 0.2 mg/ml), supplemented with defined concentrations of zinc acetate. The resulting solutions were incubated at 37°C for 30 minutes to allow zinc binding. The substrates and cofactors were then added and the activity measured spectrophotometrically.

Comet assay

The comet assay was performed essentially as described in Jugan et al. [46]. Briefly, microscope slides were coated with 1% normal melting point agarose (NMA) and allowed to dry. Around 10,000 cells (75 µL of each cell suspension) were mixed with 0.6 % low melting point agarose (LMPA) and deposited over the agarose layer, and the LMPA/cells mix was allowed to solidify on ice. The slides were immersed overnight in cold lysis solution (2.5 M NaCl, 100 mM EDTA, 10 mM Tris, 1% SDS, 10% DMSO, 1 % Triton X-100) at 4°C. DNA was then allowed to unwind for 30 min in alkaline electrophoresis solution (300 mM NaOH, 1 mM EDTA, pH > 13). Electrophoresis was performed in a field of 0.7 V/cm and 300 mA current for 30 min. Slides were then neutralized with 0.4 M Tris pH 7.5 and stained with 50 µL of 20 µg/ml ethidium bromide. At least 50 comets per slide were analyzed under a fluorescence microscope (Zeiss) equipped with a 350-390 nm excitation and a 456 nm emission filter at ×20 magnification. Comet length and intensity were measured by using Comet IV software (Perceptive Instruments, Suffolk, UK).

Proteomics

Sample preparation, 2D gel analysis and mass spectrometry were performed essentially as previously described [42] and the detailed procedures can be found in the supplementary material and methods. Thus, only the specific methods will be described in detail in this section

Sample preparation: zinc-binding proteins

Zinc-binding proteins were prepared by an affinity enrichment, using zinc ion loaded onto iminodiacetic agarose (Sigma#I4510). One milliliter of gel suspension was diluted with one milliliter of gel suspension buffer (Hepes 20 mM pH 7.5, KCl 50 mM), incubated for 10 minutes in a cold room on a rotating wheel, and centrifuged at 10,000g for 10 minutes. The gel pellet was then resuspended in 1.5 ml of gel suspension buffer, and the agitation-centrifugation cycle was repeated. The gel pellet was then resuspended in 1.5 ml of gel suspension buffer, to which zinc acetate was added at a final concentration of 0.1 M. The suspension was agitated on the rotating wheel (cold room) for 30 minutes, then centrifuged as described above. Five rinsing cycles (suspension in 1.5 ml of gel suspension buffer, rotating wheel for 15 minutes, centrifugation) were then applied. The gel was ready for use after the last centrifugation.

The native cell extract was prepared as follows: Cells were collected by scraping, centrifuged (200g for 5 minutes), and rinsed three times with PBS. The cell pellet volume was estimated, a cell lysis solution (Hepes 20 mM pH 7.5, MgCl2 2mM, KCl 50 mM, tetradecyldimethylammonio propane sulfonate (SB 3-14) 0.15 % w/v, phenylmethylsulfonyl fluoride1 mM and iodoacetamide 1 mM) was added at a ratio of 5 volumes of solution per volume of cell pellet, and the cells were left to lyse on ice for 20 minutes. The suspension was then centrifuged at 1000 g for 5 minutes, the supernatant collected, and the protein concentration determined by a dye-binding assay [65].

One milligram of proteins was loaded onto 0.2 ml of zinc-loaded iminodiacetic agarose, and the final volume was brought to 1.5 ml with gel suspension buffer. The proteins were left to adsorb overnight (rotating wheel, cold room). The gel was collected by centrifugation (10,000g 10 minutes), and six rinsing cycles (suspension in 1.5 ml of gel suspension buffer, rotating wheel for 15 minutes, centrifugation) were then applied. SB 3-14 (0.1% final concentration) was added to the rinsing buffer for the first four cycles. After the final centrifugation, four volumes (respective to the gel volume) of concentrated lysis buffer (urea 8.75 M, thiourea 2.5 M, 3-[3-Cholamidopropyl]-dimethylammoniopropane-1-sulfonate (CHAPS) 5% w/v, tris(carboxyethyl)phosphine-HCl 6.25 mM, spermine base 12.5 mM) were added and the solution was let to extract at room temperature for 1 hour. The gel was removed by centrifugation (15,000g, 20 minutes) and the protein concentration in the supernatant was determined by a dye-binding assay. Carrier ampholytes (Pharmalytes pH 3-10) were added to a final concentration of 0.4% (w/v), and the samples were kept frozen at -20°C until use.

2D gel electrophoresis

Isoelectric focusing was carried out on home made 160 mm long 4-8 linear pH gradient gels [66], cast according to published procedures [67], and rehydrated overnight with the sample in 7 M urea, 2 M thiourea, 4% CHAPS, 0.4% carrier ampholytes (Pharmalytes 3-10) and 100 mM dithiodiethanol [68,69].
IEF strips were run for 60-70 kVh (see supplementary material and methods for details), equilibrated for 20 minutes in Tris 125mM, HCl 100mM, SDS 2.5%, glycerol 30% and urea 6 M [70], and transferred on top of the SDS gels.
Ten percent gels (160x200x1.5 mm) were used for the second dimension. The Tris taurine buffer system [71] was used and operated at a ionic strength of 0.1 and a pH of 7.9. Detection was carried out by fast silver staining [72].

Image analysis

The gel images were analyzed using the Delta 2D software (v 3.6). Three gels coming from three independent cultures were used for each experimental group. Spots that were never expressed above 100 ppm of the total spots were first filtered out. Then, significantly-varying spots were selected on the basis of their Student T-test p-value between the treated and the control groups. Spots showing a p-value lower than 0.05 were selected.

Mass spectrometry

The spots selected for identification were analyzed by NanoLC-MS/MS analysis, performed using a nanoLC-QTOF-MS system and a nanoLC-IT-MS system operated as described previously [42] (details can also be found in the supplementary material).
The MS/MS data were interpreted using MASCOT 2.4.0 (Matrix Science, London, UK) against UniProtKB/SwissProt (version 2012_08, 537,505 sequences). The search was carried out in all species. A maximum of one trypsin missed cleavage was allowed.

Spectra from Qtof were searched with a mass tolerance of 15 ppm for MS and 0.05 Da for MS/MS data and spectra from Ion Trap were searched with a mass tolerance of 0.3 Da in MS and MS/MS modes. Carbamidomethylation of cysteine residues and oxidation of methionine residues were specified as variable modifications. Protein identifications were validated with at least two peptides with Mascot ion score above 20.

# AUTHOR CONTRIBUTIONS

ST performed the proteomics experiments with nanoparticles, the phagocytosis, NO experiments, and pyruvate assay. CAG also performed phagocytosis, glutathione and NO experiments, and helped in drafting the manuscript. LA performed the comet assays, and critically read the manuscript. MC helped to perform the comet assays, helped in designing the whole study and in drafting the manuscript, and critically revised the manuscript. HD, JMS and AVD performed and interpreted the mass spectrometry identification in the proteomics experiments, and helped in drafting the manuscript. FP and DH performed the transmission microscopy experiments on cells, including the EDX experiments.
AG, AH and NH performed and interpreted all the characterization experiments of the nanomaterials.
VCF performed the proteomics experiments on zinc ions and the glutathione dosage experiments. TR conceived and designed the whole study, took part in the proteomics experiments (especially metalloproteomics), performed the zinc dissolution studies, the proteasome inhibition experiments, the enzymatic measurements, and drafted the manuscript. All authors read and approved the manuscript.

ACKNOWLEDGEMENTS
ST thanks the Université Joseph Fourier for a PhD fellowship. The financial support of the CEA toxicology program (Nanobiomet and Nanostress grants) and of ANSES (Innimunotox project) is also gratefully acknowledged. Finally the support of the Labex SERENADE (11-LABX-0064) and of ARMESA is also acknowledged.

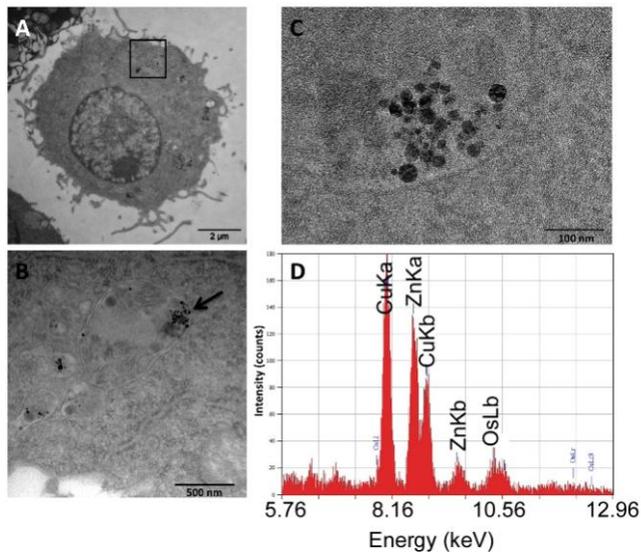

Figure 1: Transmission electron microscopy analysis of nanoparticles-treated cells

(A) TEM image of a RAW 264 cell exposed to zinc oxide. (B) A higher magnification of the boxed area in A is shown ( flipped 180°) and the arrow points to nanoparticles concentrated in multivesicular bodies. (C) Electron-dense particles resembling zinc oxide were observed in a multivesicular body by TEM and (D) the corresponding EDX spectrum, indicating the presence of zinc (arrows) which appears overlapping with a background generated from other elements present in the sample (copper for the grids and osmium for the cell post-fixation step).

A similar internalisation was also observed for the cationic coated zinc oxide nanoparticles (data not shown)

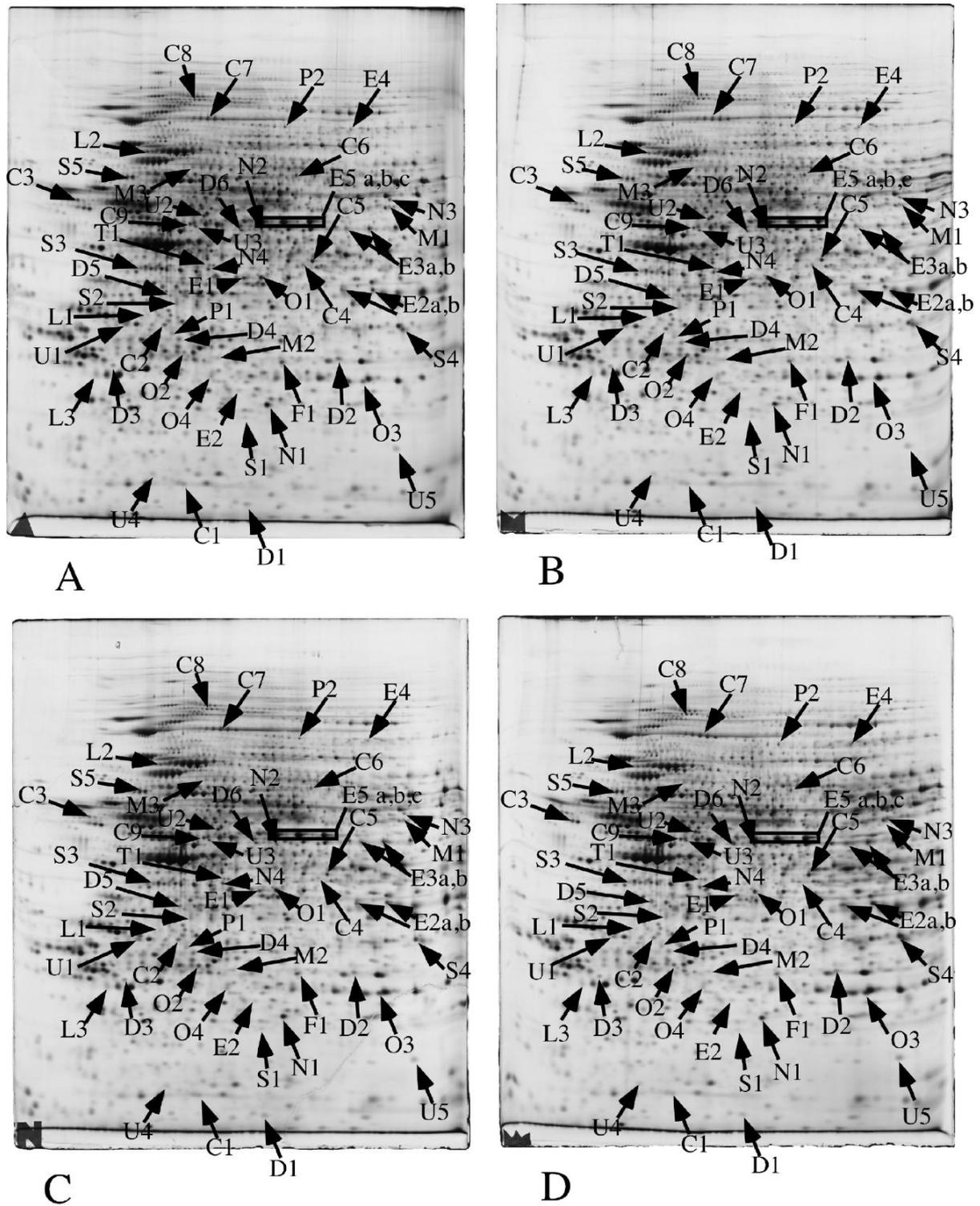

Figure 2: Proteomic analysis of total cell extracts by 2D electrophoresis

Total cell extracts of RAW264 cells were separated by two-dimensional gel electrophoresis. The first dimensions covered a 4-8 pH range and the second dimension a 15-200 kDa range. Total cellular proteins (150 µg) were loaded on the first dimension gel.

A: gel obtained from control cells
B: gel obtained from cells treated with cationic zinc oxide (7 µg/ml, 24 hours)

C: gel obtained from cells treated with zinc oxide (8 µg/ml, 24 hours)
D: gel obtained from cells treated with zinc ion (120 µM, 24 hours)

The arrows point to spots that show reproducible and statistically significant changes between the control and zinc-treated cells. Spot numbering according to Table 3.

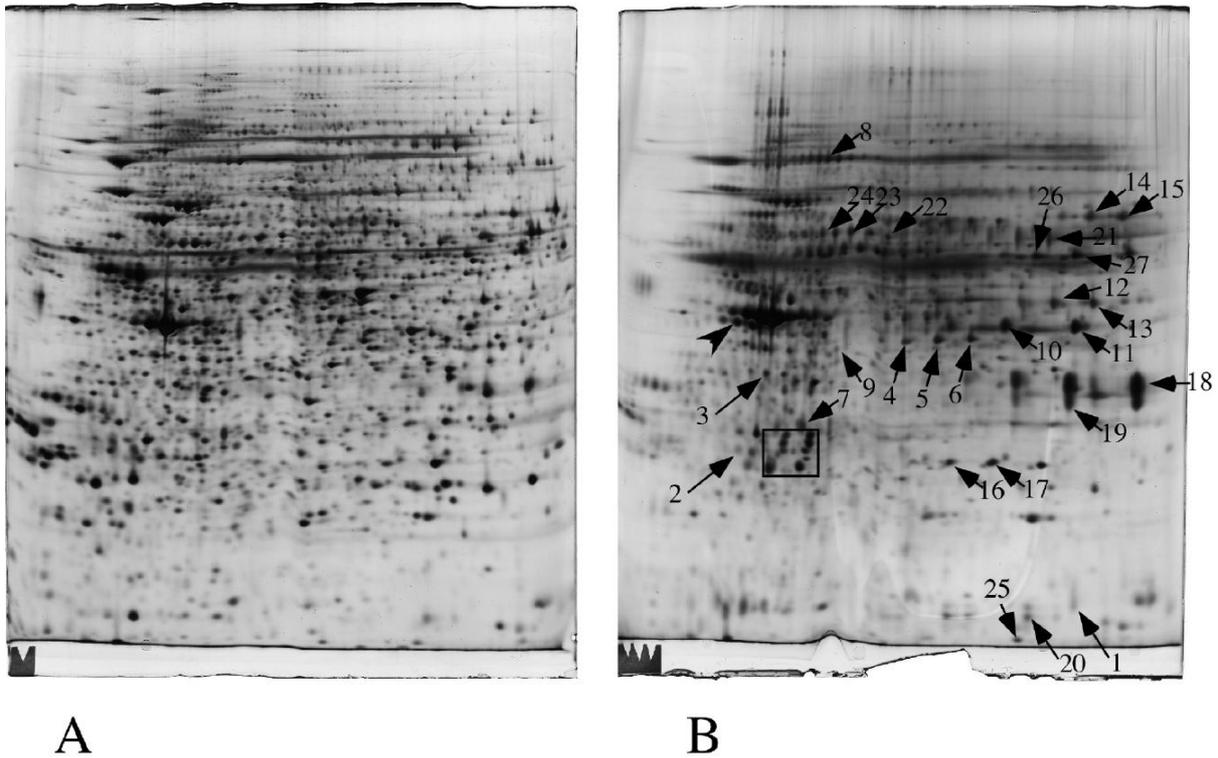

Figure 3: Proteomic analysis of zinc-binding proteins by 2D electrophoresis

Cell extracts of RAW264 cells were separated by two-dimensional gel electrophoresis. The first dimensions covered a 4-8 pH range and the second dimension a 15-200 kDa range. Two hundred micrograms of proteins were loaded on the first dimension gel.

A: analysis of the total post nuclear supernatant
B: eluate from the Zn-iminodiacetic sepharose column

The arrows point to spots that show a reproducible and statistically significant increase in the column eluate compared to the starting extract. Spot numbering according to Table 4.

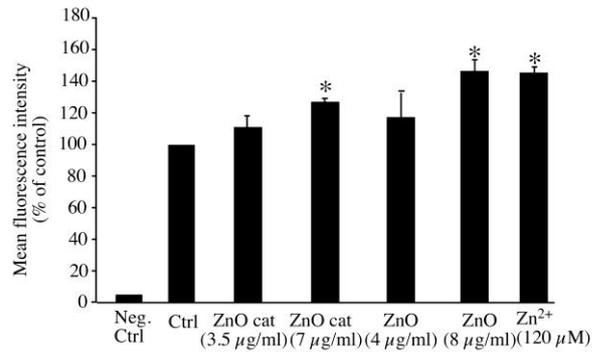

A

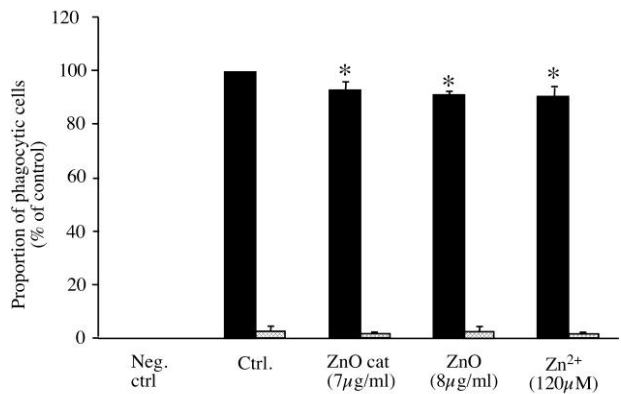

B

Figure 4: Functional alterations of macrophages upon zinc treatment

RAW264 cells were treated for 24 hours with the indicated zinc compounds. In the top panel, the intracellular concentration of reduced glutathione is indicated. The results are expressed in percentage of the activity of control cells.

In the bottom panel, the phagocytic activity index of the cells is indicated. Solid bars, activity at 37°C, hatched bars, background measurements at 4°C. The results are expressed in percentage of the activity of control cells at 37°C

Both measurements were carried out in triplicate. Statistical confidence (Student T-test) is indicated as follows: *: p ≤ 0.05; ** p ≤ 0.01; *** p ≤ 0.001

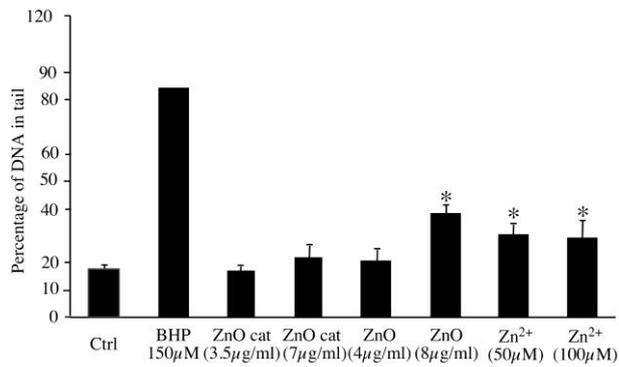

Figure 5: assessment of DNA damage
Damages to DNA on RAW264 cells after zinc treatment was evaluated using the alkaline version of the comet assay, taking into account both single and double strand breaks. The results are expressed in percentage of DNA in the tail. Butylhydroperoxide (BHP) was used as a positive control.

Measurements were carried out in triplicate (except for BHP where n=1). Statistical confidence (Student T-test) is indicated as follows: *: p ≤ 0.05; ** p ≤ 0.01; *** p≤ 0.001

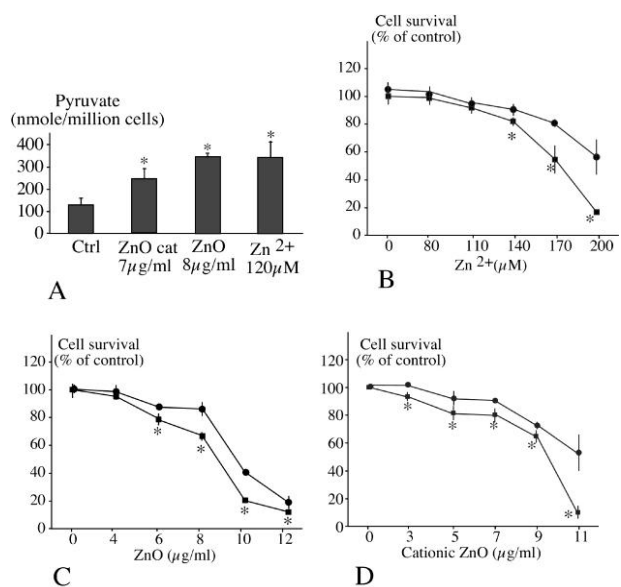

Figure 6: Role of pyruvate as a survival factor during zinc treatment
In panel A, RAW 264 cells were treated for 24 hours as indicated. The pyruvate content of the cells was measured and normalized to the number of cells after harvest.
In panels B to D, cells were pretreated with 4 mM pyruvate (circles) or not (squares) for 6 hours before treatment for a further 18 hours with zinc as indicated. Cell survival is indicated as the percentage of untreated cells. B: zinc ion; C: zinc oxide; D: cationic zinc oxide

Experiments were carried out in quadruplicate. Star denotes statistically significant differences ($p \leq 0.05$ in the Mann Whitney U test)

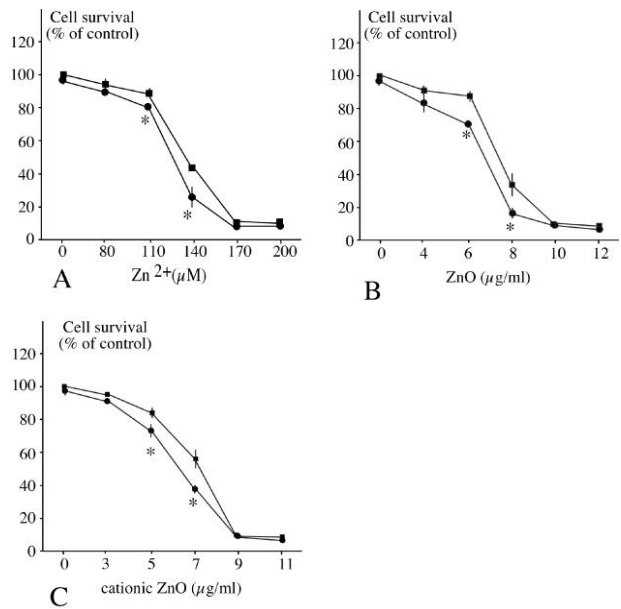

Figure 7: Role of proteasome as a survival factor during zinc treatment
RAW 264 cells were pretreated with 100 nM MG132 (circles) or not (squares) for 6 hours before treatment for a further 18 hours with zinc as indicated. Cell survival is indicated as the percentage of untreated cells. A: zinc ion; B: zinc oxide; C: cationic zinc oxide.
Experiments were carried out in quadruplicate. Star denotes statistically significant differences (p ≤ 0.05 in the Mann Whitney U test)

Table 1 : Characterization of the two zinc oxide nanoparticles used in this study

|  | ZnO | Cationic coated ZnO |
|---|---|---|
| Shape | rod | spheroid |
| size of primary particles (TEM) | 117±22nm x 49±16nm | 32±8 nm |
| zeta potential | 23.5 mV | 35 mV |
| XRD | conform | conform |
| crystaline domain size (XRD) | ca.51 nm | ca.25 nm |
| percentage in mass of organic matter (coating) | NA | 11.5% |
| average size (diameter) in water (DLS) | 245nm | 145nm |
| average size (diameter) in culture medium | 250nm | 215nm |
| Dissolved zinc in the 50mg/ml working suspension | 0.93 ± 0.12 µM | 1.19±0.37µM |

Table 2 : intracellular uptake of zinc under the different conditions

| Input in cell culture | Concentration in cell extract | Concentration in cells |
|---|---|---|
| Cationic zinc oxide (90µM) | 27±7µM | 135±35 µM |
| Zinc acetate (100µM) | 37.5±2.5 µM | 187.5±12.5 µM |
| Zinc oxide (100µM) | 30±1µM | 150±5 µM |

Table 3: Differentially-expressed proteins identified in the proteomic screen

| spot ID | protein name | accession number* | ZnOcat /ctrl fold/T-test | ZnO /ctrl fold/T-test | Zn ion/ctrl fold/T-test | nb. of unique peptides | sequence coverage |
|---|---|---|---|---|---|---|---|
| F: folding | | | | | | | |
| F1 | Endoplasmic reticulum resident protein 29 | P57759 | 1.003/98 | 1.008/92 | 0.808/4 | 6 | 24% |
| F2 | Prefoldin 3 | P61759 | 1.111/59 | 0.599/9 | 0.645/8.4 | 3 | 21% |
| C: cytoskeleton | | | | | | | |
| C1 | Glia maturation factor gamma | Q9ERL7 | 0.753/19 | 0.486/18 | 0.620/7.4 | 5 | 39% |
| C2 | Microtubule-associated protein MPREB1 | Q61166 | 0.969/73 | 0.872/14 | 0.809/3.4 | 7 | 40% |
| C3 | Lymphocyte-specific protein 1 | P19973 | 0.815/10 | 0.824/5 | 0.755/3 | 5 | 21% |
| C4 | Twinfilin-2 | Q9Z0P5 | 0.835/36 | 0.832/36 | 0.755/3 | 6 | 22% |
| C5 | Actin-related protein 2 | P61161 | 0.855/8 | 0.841/12 | 0.862/9 | 10 | 29% |
| C6 | T-complex protein 1 subunit gamma | P80318 | 1.022/84 | 0.988/99 | 0.618/2 | 6 | 12% |
| C7a | Gelsolin | P13020 | 1.26/41 | 1.50/5 | 1.28/39 | 6 | 8% |
| C7b | Gelsolin | P13020 | 1.22/44 | 1.51/6 | 1.22/41 | 4 | 8% |
| C7c | Gelsolin | P13020 | 1.28/31 | 1.65/2 | 1.35/22 | 5 | 7% |
| C7d | Gelsolin | P13020 | 1.45/37.2 | 1.876/6 | 1.569/18 | 11 | 17% |
| C7e | Gelsolin | P13020 | 1.72/25 | 2.3/4 | 17/18 | 6 | 8% |
| C8 | Major vault protein | Q9EQK5 | 1.643/4.1 | 1.599/7 | 1.469/10 | 12 | 17% |
| C9 | Actin-related protein 3 | Q99JY9 | 1.23/3 | 1.19/11 | 1.25/18 | 5 | 15% |
| D: degradation | | | | | | | |
| D1 | **Ubiquitin-conjugating enzyme E2 N** | P61089 | 0.734/19 | 0.600/3 | 0.741/1.3 | 3 | 27% |
| D2 | **Calcyclin-binding protein** | Q9CXW3 | 1.111/67 | 1.072/22.4 | 0.862/5 | 7 | 30% |
| D3 | **Ubiquitin carboxyl-terminal hydrolase L3** | Q9JKB1 | 0.667/7 | 1.024/88 | 0.58/10 | 3 | 14% |
| D4 | **Proteasome activator complex subunit 2** | P97372 | 0.943/33 | 0.98/86 | 0.832/1.3 | 6 | 37% |
| D5 | **Ubiquitin carboxyl-terminal hydrolase L5** | Q9WUP7 | 0.972/73 | 0.908/8.4 | 0.495/2 | 8 | 32% |
| D6 | **26S protease regulatory subunit 7** | P46471 | 0.761/13 | 0.762/11.3 | 0.518/2.2 | 3 | 8% |
| S: signalling | | | | | | | |
| S1 | Growth factor receptor-bound protein 2 | Q60631 | 0.622/14.5 | 0.666/3 | 0.466/2 | 4 | 16% |
| S2 | G nucleotide-binding protein beta-2 | P62880 | 0.866/34 | 0.892/13 | 0.638/1.6 | 4 | 13% |
| S3 | G nucleotide-binding protein alpha-2 | P08752 | 0.878/19 | 0.848/12 | 0.712/2.4 | 5 | 21% |
| S4 | GNB protein subunit beta-2-like 1 | P68040 | 1.719/3.2 | 1.674/7.9 | 1.54/8.1 | 11 | 42% |
| S5 | PP2A 2A 65 kDa scaffold subunit A alpha | Q76MZ3 | 1.087/52 | 1.246/4 | 1.177/15 | 5 | 19% |
| N: nucleotide metabolism | | | | | | | |
| N1 | Adenine phosphoribosyltransferase | P47957 | 0.86/7 | 0.918/2 | 0.943/54 | 6 | 54% |

| | | | | | | |
|---|---|---|---|---|---|---|
| N2 | *Adenylosuccinate synthetase isozyme 2* | P46664 | 0.761/13 | 0.762/3 | 0.752/18 | 4 | 12% |
| N3 | IMP DH 2 | P24547 | 0.676/1 | 0.784/3 | 0.752/12 | 3 | 7% |
| N4 | 3'(2')5'-bisphosphate nucleotidase 1 | Q9Z0S1 | 1.262/48 | 1.042/85 | 0.439/5 | 7 | 29% |

E: energy metabolism

| | | | | | | |
|---|---|---|---|---|---|---|
| E1 | Transaldolase | Q93092 | 0.912/54 | 1.047/73 | 0.842/48 | 6 | 20% |
| E2a | Aldose reductase | P45376 | 1.054/57 | 1.206/2 | 1.306/9 | 7 | 28% |
| E2b | Aldose reductase | P45376 | 1.25/17 | 1.42/11 | 1.95/1 | 7 | 26% |
| E3a | **6-phosphogluconateDH, decarboxylating** | Q9DCD0 | 1.135/47 | 1.258/7.7 | 1.305/3.8 | 8 | 20% |
| E3b | **6-phosphogluconateDH, decarboxylating** | Q9DCD0 | 1.08/60 | 1.33/5 | 1.25/17 | 14 | 35% |
| E4 | 6-phosphofructokinase type C | Q9WUA3 | 0.473/4 | 1.19/31 | 1.29/26 | 9 | 16% |
| E5a | **enolase** | P17182 | 1.23/1 | 1.33/4 | 1.20/11 | 17 | 57% |
| E5b | **enolase** | P17182 | 1.23/23 | 1.33/9 | 1.33/12 | 23 | 70% |
| E5c | **enolase** | P17182 | 1.27/35 | 1.43/15 | 1.5/14 | 24 | 67% |

O: oxidative stress response

| | | | | | | |
|---|---|---|---|---|---|---|
| O1 | Neutrophil cytosol factor 4 | P97369 | 0.991/95 | 1.061/72 | 0.53/3.3 | 7 | 28% |
| O2 | **Glutamate-cysteine ligase reg.subunit** | O09172 | 1.631/7 | 1.644/2 | 1.837/3 | 5 | 22% |
| O3 | **Flavin/biliverdin reductase (NADPH)** | Q923D2 | 1.571/7 | 1.126/46 | 1.467/7 | 6 | 39% |
| O4 | Peroxiredoxin-6 | O08709 | 0.832/3 | 1.066/59 | 0.792/4 | 4 | 26% |

M: mitochondrion

| | | | | | | |
|---|---|---|---|---|---|---|
| M1 | Glutathione reductase. mitochondrial | P47791 | 2.153/1 | 2.507/2 | 2.157/1 | 2 | 4% |
| M2 | Putative Clp protease. mitochondrial | O88696 | 1.408/8.6 | 1.342/4.2 | 1.007/95 | 4 | 18% |
| M3 | *DLAT. mitochondrial* | Q8BMF4 | 0.566/7.7 | 1.046/76 | 1.2/6 | 2 | 5% |

P: proliferation

| | | | | | | |
|---|---|---|---|---|---|---|
| P1 | Replication protein A 32 kDa subunit | Q62193 | 0.797/14 | 0.882/45 | 0.764/10 | 3 | 13% |
| P2 | DNA replication licensing factor MCM7 | Q61881 | 1.45/2 | 1.17/11 | 1.13/50 | 15 | 28% |

T: translation

| | | | | | | |
|---|---|---|---|---|---|---|
| T1 | Methylthioribose-1-phosphate isomerase | Q9CQT1 | 1.097/31 | 0.823/23 | 0.462/0.7 | 5 | 17% |

L: lysosome

| | | | | | | |
|---|---|---|---|---|---|---|
| L1 | Alpha-soluble NSF attachment protein | Q9DB05 | 1.017/93 | 1.060/34 | 0.711/3 | 4 | 15% |
| L2 | Sorting nexin-2 | Q9CWK8 | 0.843/10 | 0.901/42 | 0.729/2.1 | 11 | 25% |
| L3 | Synaptosomal-associated protein 23 | O09044 | 0.613/4 | 0.580/7 | 0.464/1.5 | 3 | 22% |
| U: Miscellaneous | | | | | | | |
| U1 | EF-hand domain-containing protein D2 | Q4FZY0 | 0.904/33 | 0.88/26 | 0.716/3.7 | 6 | 28% |
| U2 | RuvB-like 2 | Q9WTM5 | 1.027/75 | 0.845/2 | 0.863/17 | 11 | 26% |
| U3 | Dolichyl-diphosphooligosaccharide--protein glycosyltransferase 48 kDa subunit | O54734 | 1.39/3 | 1.27/3 | 1.427/23 | 6 | 15% |
| U4 | Glia maturation factor beta | Q9CQI3 | 0.824/31 | 0.692/6.4 | 0.755/4.2 | 4 | 31% |
| U5 | Transcription factor BTF-3 | Q64152 | 1.44/28 | 1.34/21 | 1.57/5 | 4 | 38% |

\*The accession numbers are those of the SwissProt Database. The proteins mentioned in the text are italicized in the table. The proteins validated directly or indirectly are in bold in the table.

Table**4**: Zinc-binding proteins identified in the metalloproteomic screen

| spot ID | protein name | accession number* | Mw | IDA/PNS (fold/T-test) | Nb. unique peptides | sequence coverage |
|---|---|---|---|---|---|---|
| Traffic and cytoskeleton | | | | | | |
| 1 | Actin-related protein 2/3 complex subunit 4 | P59999 | 19667.4 | 5.6/0 | 3 | 18% |
| 2 | Ran-specific GTPase-activating protein | P34022 | 23597.1 | 1.28/10 | 4 | 21% |
| 3 | F-actin-capping protein subunit alpha-1 | P47753 | 32939.9 | 2.18/0.1 | 12 | 49% |
| 4 | Actin-related protein 2 | P61161 | 44761.7 | 5.1/0.2 | 14 | 36% |
| 5 | Actin-related protein 2 | P61161 | 44761.7 | 1.6/4.6 | 19 | 50% |
| 6 | Actin-related protein 2 | P61161 | 44761.7 | 1.04/73 | 12 | 32% |
| 7 | F-actin-capping protein subunit beta | P47757 | 33742.2 | 1.56/0.3 | 5 | 18% |
| 8 | Gelsolin | P13020 | 85941.9 | 1.77/0.4 | 37 | 47% |
| Energy metabolism | | | | | | |
| 9 | Isocitrate dehydrogenase [NAD] subunit alpha | Q9D6R2 | 39639.4 | 1.97/10 | 8 | 25% |
| 10 | Phosphoglycerate kinase 1 | P09411 | 44551.1 | 3.45/0.2 | 28 | 64% |
| 11 | Phosphoglycerate kinase 1 | P09411 | 44551.1 | 2.23/0.4 | 34 | 77% |
| 12 | 6-phosphogluconate dehydrogenase | Q9DCD0 | 53248.8 | 1.86/0.3 | 22 | 38% |
| 13 | 6-phosphogluconate dehydrogenase | Q9DCD0 | 53248.8 | 1.34/0.7 | 23 | 43% |
| 14 | Transketolase | P40142 | 67631.9 | 3.36/0.8 | 24 | 37% |
| 15 | Transketolase | P40142 | 67631.9 | 3.29/0.3 | 37 | 55% |
| 16 | Phosphoglycerate mutase 1 | Q9DBJ1 | 28804.8 | 3.39/0 | 12 | 43% |
| 17 | Phosphoglycerate mutase 1 | Q9DBJ1 | 28804.8 | 2.04/0.1 | 13 | 48% |
| 18 | Glyceraldehyde-3-phosphate dehydrogenase | P16858 | 35810.1 | 11.5/0 | 14 | 43% |
| 19 | Glyceraldehyde-3-phosphate dehydrogenase | P16858 | 35810.1 | 2.72/7 | 9 | 31% |
| Folding and chaperones | | | | | | |
| 20 | Peptidyl-prolyl cis-trans isomerase A | P17742 | 17971.8 | 1.47/2.5 | 15 | 74% |
| 21 | T-complex protein 1 subunit zeta | P80317 | 58005.5 | 1.27/0.6 | 26 | 56% |
| 22 | T-complex protein 1 subunit alpha | P11983 | 60449.9 | 1.26/3.2 | 15 | 30% |
| 23 | T-complex protein 1 subunit alpha | P11983 | 60449.9 | 2.43/0.6 | 24 | 47% |
| 24 | T-complex protein 1 subunit alpha | P11983 | 60449.9 | 2.30/1.5 | 16 | 36% |
| 24 | T-complex protein 1 subunit epsilon | P80316 | 59625.6 | | 16 | 34% |
| Miscellaneous | | | | | | |

| 25 | Nucleoside diphosphate kinase A | P15532 | 17208 | 1.84/4.5 | 14 | 76% |
| 26 | Adenylyl cyclase-associated protein 1 | P40124 | 51565.1 | 1.20/34 | 29 | 63% |
| 27 | Adenylyl cyclase-associated protein 1 | P40124 | 51565.1 | 1.74/0.2 | 31 | 72% |

*The accession numbers are those of the SwissProt Database.

Table 5 : enzyme acitivities measured in control cell extracts and in extracts prepared from cells treated for 24 hours with either 8µg/ml zinc oxide, or 7µg/ml cationic zinc oxide, or 100µM zinc acetate.
The activities are expressed in units/mg protein, the unit being defined as 1µmole of substrate converted per minute

| enzyme | enolase | 6PGDH | biliverdin reductase |
|---|---|---|---|
| ctrl | 55.3±4.2 | 50.1±6.3 | 1.53±0.16 |
| Cationic ZnO | 84.15±8.9 | 59.7±4.6 | 2.05±0.02 |
| ZnO | 98±9.7 | 94.6±0.2 | 2.01±0.14 |
| Zn acetate | 87±2.2 | 92.7±4.0 | 2.17±0.09 |
|  |  |  |  |
|  | 6PGDH |  |  |
| 0µM Zn | 50.5±5.5 |  |  |
| 10µM Zn | 33±2 |  |  |
| 15µM Zn | 20±1 |  |  |
| 20µM Zn | 9.5±0.75 |  |  |
| 25µM Zn | 4.75±0.5 |  |  |